# The Cosmic Digital Code and Quantum Mechanics

Ding-Yu Chung*

Resembling two-value digital code for computer, the cosmic digital code as the law of all physical laws contains two mutually exclusive values: attachment space attaching to object and detachment space detaching from object.  However, the cosmic physical system could not start with mutually exclusive attachment space and detachment space at the same time in the beginning.  The way out of this impasse is that the complete cosmic system consists of both the cosmic physical system and the unphysical cosmic digital code.  The unphysical cosmic digital code as the law of all physical laws allows the coexistence of attachment space and detachment space at the same time.   The cosmic digital code behaves as gene in organism, and the cosmic physical system behaves as organ.  Under different conditions and times, different spaces in the cosmic digital code are activated to generate different spaces in the cosmic physical systems with different physical laws for different universes in the multivese.  All universes start from the primitive multiverse, which only has attachment space attaching to 10D string without the four force fields.  During the big bang in our universe, detachment space emerged to detach particle from its four force fields.  Such detachment, however, cannot be completely and permanently detached, because particle still attaches to its force fields.  The result is hybrid space, combining both attachment space and detachment space in coherent state, such as particle and its force fields.  Hybrid space is the space for wavefunction whose probability density is proportional to attachment space, and inversely proportional to detachment space.  The collapse of wavefunction in decoherent state is the separation of attachment space and detachment space in such way that attachment space attaches to particle, and detachment space separately detaches from all probability density.

## Introduction

Two-value digital code for computer contains two mutually exclusive values: 1 and 0, representing on and off.  This paper posits that the cosmic digital code as the law of all physical laws also contains two mutually exclusive values: attachment space attaching to object and detachment space detaching from object. However, the cosmic physical system could not start with mutually exclusive attachment space and detachment space at the same time in the beginning.  At the start of the cosmic physical system, when attachment space attaching to object appeared, detachment space detaching from object had to appear outside of the cosmic physical system.  Attachment space and detachment space could not be both coexisted and complete at the start of the cosmic physical system.  The way out of this impasse is that the complete cosmic system consists of both the cosmic physical system and the unphysical cosmic digital code.  The unphysical cosmic digital code as the law of all physical laws allows the coexistence of attachment space and detachment space. The cosmic digital code behaves as gene in organism, and the cosmic physical system behaves as organ. Under different conditions and times, different spaces in the cosmic digital code are activated to generate different spaces in the cosmic physical system with different physical laws for different universes in the multivese. This paper describes the cosmic process for the activation of attachment space and detachment space at different times for our universe, and relates such cosmic process to quantum mechanics.

## The Varying Dimension Number

As described previously [1], all universes start from the primitive multiverse, which has attachment space attaching to 10D (space-time dimensional) string without the four force fields. There is no detachment space in the primitive multiverse. Detachment space emerged later during the big bang to detach particle from its force fields. In this paper, the force fields are derived from the extra space-time dimensions due to varying dimension numbers [1].

Varying dimension numbers are derived from varying speed of light (VSL) theory [2]. The constancy of the speed of light is the pillar of special relativity. The constancy of the speed of light takes place in the four dimensional space-time whose space-time dimension number (four) is constant. In the varying speed of light (VSL) model [3] of cosmology by Albrecht, Magueijo, and Barrow, the speed of light varies in time. The time dependent speed of light varies as some power of the expansion scale factor, a, in such way that

$$c(t) = c_0 \, a^n \tag{1}$$

where $c_0 > 0$ and n are constant. The increase of speed of light is continuous.

This paper posits quantized varying speed of light (QVSL), where the speed of light is invariant in a constant space-time dimension number, and the speed of light varies with varying space-time dimension number from 4 to 11. In QVSL, the speed of light is quantized by varying space-time dimension number.

$$c_D = c / \alpha^{D-4}, \tag{2}$$

where c is the observed speed of light in the 4D space-time, $c_D$ is the quantized varying speed of light in space-time dimension number, D, from 4 to 11, and $\alpha$ is the fine structure constant. Each dimensional space-time has a specific speed of light. The speed of light increases with increasing space-time dimension number, D. In the VDN model of cosmology, the universe starts with the pre-expanding universe that has the speed of light in 11D space-time.

In special relativity, $E = M_0 c^2$ modified by eq. (2) is expressed as

$$E = M_0 (c^2 / \alpha^{2(D-4)}) \tag{3a}$$
$$= (M_0 / \alpha^{2(d-4)}) \, c^2 \tag{3b}$$

Eq. (3a) means that a particle in the D dimensional space-time can have superluminal speed, $c / \alpha^{D-4}$, that is higher than the observed speed of light, and has rest mass, $M_0$. Eq. (3b) means that the same particle in the 4D space-time with the observed speed of light acquires $M_0/\alpha^{2(d-4)}$ as the rest mass where d = D. D in eq. (3a) is space-time dimension number defining the varying speed of light. In eq. (3b), d from 4 to 11 is "mass dimension number" defining varying mass. For example, for D = 11, eq. (3a) shows a superluminal particle in eleven-dimensional space-time, while eq. (3b) shows that the speed of light of the same particle is the observed speed of light with the 4D space-time, and the mass dimension is eleven. In other words, 11D space-time can transform into 4D space-time with 11d mass dimension. QVSL in terms of varying space-time dimension number, D, brings about varying mass in terms of varying mass dimension number, d.

The QVSL transformation transforms space-time dimension number and mass dimension number. In the QVSL transformation, the decrease in the speed of light leads to the decrease in space-time dimension number and the increase of mass in terms of increasing mass dimension number from 4 to 11.

$$c_D = c_{D-n} / \alpha^{2n}, \tag{4a}$$

$$M_{0,D,d} = M_{0,D-n,\ d+n} \alpha^{2n}, \tag{4b}$$

$$D, d \xrightarrow{QVSL} (D \mp n),\ (d \pm n) \tag{4c}$$

where D is space-time dimension number from 4 to 11 and d is mass dimension number from 4 to 11. For example, the QVSL transformation transforms a particle with 11D4d to a particle with 4D11d. In terms of rest mass, 11D space-time has 4d with the lowest rest mass, and 4D space-time has 11d with the highest rest mass.

In the normal supersymmetry transformation, the repeated application of the fermion-boson transformation transforms a boson (or fermion) from one point to the same boson (or fermion) at another point at the same mass. In the "varying supersymmetry transformation", the repeated application of the fermion-boson transformation transforms a boson from one point to the boson at another point at different mass dimension number in the same space-time number. The repeated varying supersymmetry transformation transforms boson $B_d$ into fermion $F_d$ and from fermion $F_d$ to boson $B_{d-1}$ is expressed as

$$M_{d,F} = M_{d,B}\ \alpha_{d,B}, \tag{5a}$$

$$M_{d-1,B} = M_{d,F}\ \alpha_{d,F}, \tag{5b}$$

where $M_{d,B}$ and $M_{d,F}$ are the masses for a boson and a fermion, respectively, d is mass dimension number, and $\alpha_{d,B}$ or $\alpha_{d,F}$ is the fine structure constant, which is the ratio between the masses of a boson and its fermionic partner. Assuming $\alpha_{d,B} = \alpha_{d,F}$, the relation between the bosons in the adjacent dimensions, then, can be expressed as

$$M_{d-1,B} = M_{d,B}\ \alpha_d^2, \tag{5c}$$

Eq. 5 shows that it is possible to describe mass dimensions > 4 in terms of

$$F_5 B_5 \ F_6 B_6 \ F_7 B_7 \ F_8 B_8 \ F_9 B_9 \ F_{10} B_{10} \ F_{11} B_{11} \quad , \tag{6}$$

where the energy of $B_{11}$ is Planck energy. Each mass dimension between 4d and 11d consists of a boson and a fermion. Eq. 5 shows a stepwise transformation that transforms a particle with d mass dimension to d ± 1 mass dimension. The transformation from higher dimensional particle to adjacent lower dimensional particle is the fractionalization of a higher dimensional particle to many lower dimensional particle in such way the number of lower dimensional particles = $n_{d-1}$ = $n_d / \alpha^2$. The transformation from lower dimensional particles to higher dimensional particle is condensation. Both the fractionalization and the condensation are stepwise. For example, a particle with 4D (space-time)10d (mass dimension) can transform stepwise into 4D9d particles. Since supersymmetry transformation involves translation, this stepwise varying supersymmetry transformation leads to translational fractionalization and translational condensation, resulting in expansion and contraction.

Another type of the varying supersymmetry transformation is not stepwise. It is the leaping varying supersymmetry transformation that transforms a particle with d mass dimension to any d ± n mass dimension. The transformation involves the fission-fusion of particle. The transformation from d to d – n involves the fission of a particle with d mass dimension into two parts: the core particle with d – n dimension and the n dimensions that are separable from the core particle. Such n dimensions are denoted as n "dimensional orbitals", which become force fields, including the four force fields [1]. The sum of the number of mass dimensions for a particle and the number of dimensional orbitals is equal to 11 for all particles with mass dimensions. Therefore,

$$F_d = F_{d-n} + (11 - d + n) DO's \ , \tag{7}$$

where 11- d + n is the number of dimensional orbitals (DO's) for $F_{d-n}$. For example, the fission of 4D9d particle produces 4D4d particle that has d = 4 core particle and 7 separable dimensional orbitals in the form of $B_5 F_5 B_6 F_6 B_7 F_7 B_8 F_8 B_9 F_9 B_{10} F_{10} B_{11}$. Since the fission process is not stepwise from higher mass dimension to lower mass dimension, it is possible to have simultaneous fission. For example, 4D9d particles can simultaneously transform into 4D8d, 4D7d, 4D6d, 4D5d, and 4D4d particles, which have 3, 4, 5, 6, and 7 separable dimensional orbitals, respectively. Therefore, varying supersymmetry transformation can be stepwise or leaping. Stepwise supersymmetry transformation is translational fractionalization and condensation, resulting in stepwise expansion and contraction. Leaping supersymmetry transformation is not translational, and it is fission and fusion, resulting possibly in simultaneous formation of different particles with separable dimensional orbitals.

In summary, the QVSL transformation transform space-time dimension number and mass dimension number. The varying supersymmetry transforms varying mass dimension number in the same space-time number as follows (D = space-time dimension number and d = mass dimension number).

$$D, d \xrightarrow{QVSL} (D \mp n), (d \pm n)$$

$$D, d \xrightarrow{stepwise\ or\ leaping\ varying\ supersymmetry} D, (d \pm 1)\ or\ D, (d \pm n)$$

## Quantum Mechanics

Before the inflation, the universe is made of strings as 10D4d with another dimension for gravity. 10D4d string transforms through the QVSL transformation quickly into 4D10d particles, which then transforms and fractionalizes quickly through varying supersymmetry transformation into 4D9d, resulting in inflationary expansion [4]. The inflationary expansion occurs between the energy for 4D10d = $E_{Planck}\alpha^2 = 6 \times 10^{14}$ GeV and the energy for 4D9d = $E_{10}\alpha^2 = 3 \times 10^{10}$ GeV. At the end of the inflationary expansion, all 4D9d particles undergo simultaneous fission to generate equally by mass and number into 4D9d, 4D8d, 4D7d, 4D6d, 4D5d, and 4D4d particles. Baryonic matter is 4D4d, while dark matter consists of the other five types of particles. The mass ratio of dark matter to baryonic matter is 5 to 1 in agreement with the observation [5] that shows that the universe consists of 25% dark matter, 5% baryonic matter, and 70% dark energy. Afterward, thermal expansion (the big bang) takes place. In summary, the process is as follows.

$$10D4d \xrightarrow{QVSL\ transformation} 4D10d \xrightarrow{stepwise\ fractionalization,\ inflation}$$
$$4D9d \xrightarrow{simultaneous\ fission} 4D9d + 4D8d + 4D7d + 4D6d + 4D5d + 4D4d + radiation$$
$$\rightarrow thermal\ cosmic\ expansion\ (the\ big\ bang)$$

The mechanism for the fission into core particle and dimensional orbital requires detachment space that detaches core particle and its dimensional orbitals, which become force fields. The absorption of detachment space in the gap between core particle and dimensional orbital initiates the fission [1]. However, the gap between core particle and dimensional orbital is not purely detachment space that creates permanent detachment. The space between core particle and dimensional orbital is "hybrid space" from combining attachment space and detachment space. Hybrid space has neither complete attachment nor complete detachment. Hybrid space can be described by the uncertainty principle in quantum mechanics as

$$\Delta x \Delta p \geq h/4\pi, \tag{8}$$

where x is position and p is momentum. For the uncertainty principle in quantum mechanics, $\Delta x$ is the uncertainty in position and $\Delta p$ is the uncertainty in momentum during the measurement of the position and momentum of a particle. For hybrid space, $\Delta x$ is distance of the gap between core particle and dimensional orbital, and $\Delta p$ is the momentum in the gap from the by the contact between dimensional orbital and core particle. $\Delta x \Delta p \geq h/4\pi$ shows that neither $\Delta x$ nor $\Delta p$ can be zero. Thus, there cannot be complete attachment where $\Delta x$ = zero, and there cannot be complete detachment where $\Delta p = 0$. For hybrid space, the uncertainty principle without precise position and momentum becomes the gap principle without complete attachment and detachment.

All particles have core particle and dimensional orbital, so hybrid space applies all particles. The degree of detachment space in hybrid space determines the probability density in a wavefunction of a particle. The high degree of detachment space has the low probability density. Particle becomes particle-wave. The original purpose of hybrid space is for the gap in a coherent system of core particle and dimensional orbital, so hybrid space is coherent. Thus, hybrid space as the space for all particles is coherent. As in the theory of decoherence, any decoherence by the entanglement with decoherent system obliterates hybrid space, resulting in the collapse of wavefunction. The collapse of wavefunction in decoherent state is the separation of attachment space and detachment space in such way that attachment space attaches to particle, and detachment space separately detaches from all probability density.

In summary, the unphysical cosmic digital code contains two mutually exclusive values: attachment space attaching to object and detachment space detaching from object. Complete system consists of both the cosmic physical system and the unphysical cosmic digital code. All universes start from the primitive multiverse, which has attachment space attaching to 10D string without detachment space and the four force fields. Detachment space emerged later during the big bang to detach particle from its force fields. Such detachment, however, cannot be completely and permanently detached, because particle still attaches to its force fields. The result is hybrid space. Unlike the two-value digital code for computer, the cosmic digital code allows the third value: hybrid space that is the combination of both attachment space and detachment space in coherent state, such as particle and its force fields. Hybrid space is the space for wavefunction. In wavefunction, probability density is proportional to attachment space, and inversely proportional to detachment space. Hybrid space in wavefunction can both attach to and detach from object. The collapse of wavefunction in decoherent state is the separation of attachment space and detachment space in such way that attachment space attaches to particle, and detachment space separately detaches from all probability density.

## References

\*      chung@wayne.edu P.O. Box 180661, Utica, Michigan 48318, USA

[1]    D. Chung, hep-th/0201115, D. Chung, Speculations in Science and Technology 20 (1997) 259; Speculations in Science and Technology 21(1999) 277
[2]    G. Amelino-Camelia, Int. J. Mod. Phys. D11 (2002) 35 [gr-qc/0012051]; Phys. Letts. B510 (2001) 255 [hep-th/0012238]; J Barrow, gr-qc/0211074; G. Ellis and J. Uzan, gr-qc/0305099; J. Magueijo, astro-ph/0305457

[3] A. Albrecht and J. Magueijo, Phys. Rev. D59, 043516 [astro-ph/9811018]; J. D. Barrow, Phys. Rev. D59, 043515 (1999); J. D. Barrow, Phys.Lett. B564, 1 (2003) [gr-qc/0211074]
[4] A. H. Guth, Phys. Rev. D 23, 347 (1981), A. D. Linde, Phys. Lett. B108 (1982) 389; A. Albrecht and P. J. Steinhardt, Phys. Rev. Lett. 48 (1982) 1220
[5] M. Rees, Phil.Trans.Roy.Soc.Lond. 361 (2003) 2427